\documentclass[a4paper]{article}

\usepackage{INTERSPEECH2019}
\usepackage{multirow}
\usepackage{multicol}

\pdfoutput=1

\usepackage{comment}
\usepackage{textcomp}
\usepackage{url}

\usepackage{soul} 
\usepackage{color}

\title{Ensemble Models for Spoofing Detection in Automatic Speaker Verification}

\name{Bhusan Chettri$^1$, Daniel Stoller$^1$, Veronica Morfi$^1$, Marco A. Mart\'{i}nez Ram\'{i}rez$^1$, \\Emmanouil Benetos$^1$\thanks{EB is supported by RAEng Research Fellowship RF/128 and a Turing Fellowship. DS is funded by EPSRC grant EP/L01632X/1. This research was supported by an NVIDIA GPU Grant.}, Bob L. Sturm$^{2}$}
\address{
  $^1$School of EECS, Queen Mary University of London, United Kingdom  \\  
  $^2$School of EECS, KTH Royal Institute of Engineering, Stockholm, Sweden}
\email{}

\begin{document}
\maketitle
\begin{abstract}
Detecting spoofing attempts of automatic speaker verification (ASV) systems is challenging, especially when using only one modelling approach. For robustness, we use both deep neural networks and traditional machine learning models and combine them as ensemble models through logistic regression. They are trained to detect logical access (LA) and physical access (PA) attacks on the dataset released as part of the ASV Spoofing and Countermeasures Challenge 2019. We propose dataset partitions that ensure different attack types are present during training and validation to improve system robustness. Our ensemble model outperforms all our single models and the baselines from the challenge for both attack types. We investigate why some models on the PA dataset strongly outperform others and find that spoofed recordings in the dataset tend to have longer silences at the end than genuine ones. By removing them, the PA task becomes much more challenging, with the tandem detection cost function (t-DCF) of our best single model rising from 0.1672 to 0.5018 and equal error rate (EER) increasing from 5.98\% to 19.8\% on the development set. 

\end{abstract}

\section{Introduction}
An automatic speaker verification (ASV) \cite{sid_using_GMM} system aims at verifying the claimed identity of a speaker and is widely used for person authentication. Though the technology has matured immensely over the past few years, studies \cite{wu2013voice,wu2015_surveyspoofing} have confirmed its vulnerability in the face of spoofing, also known as a presentation attack \cite{iso_spoofing_standards}. Mimicry \cite{wah2004mimicry}, replay \cite{replay_wu2014study}, text-to-speech (TTS) \cite{wu2015_surveyspoofing} and voice-conversion (VC) \cite{wu2013voice} technology are commonly used to perform logical access (LA) or physical access (PA) spoofing attacks in ASV systems \cite{asvspoof2019overview}. While LA attacks (TTS and VC) are mounted by injecting synthetic/converted speech directly into the ASV pipeline bypassing its microphone, PA attacks (replay and mimicry), on the contrary, involve physical transmission of impersonated or playback speech through the systems' microphone. 

Spoofing countermeasures for reliable speaker verification are therefore of paramount interest. To this end, the ASV community has released standard spoofing datasets \cite{asv2015overview,asv2017overview,asvspoof2019overview} as part of the automatic speaker verification spoofing and countermeasures challenges (ASVspoof), promoting research in this direction. The ASVspoof 2019 challenge \cite{asvspoof2019_evaluationplan,asvspoof2019overview} combines both LA and PA (excluding mimicry) attacks using the latest state-of-the-art TTS and VC methods and controlled-simulation setup for replay attacks, in contrast to the 2015 and 2017 spoofing datasets. 


Designing a single model to robustly detect unseen spoofing attacks can be challenging, as demonstrated at the ASVspoof 2015 and 2017 challenges, where the best performing systems \cite{bestSystemIS2015challenge,secondbestSystemIS2015challenge,lavrentyeva2017audio,secondbestsystem_2017challenge,thirdbestsystem_2017challenge,fourthbestsystem_2017challenge} made use of an ensemble model combining features or scores. In this paper, we investigate LA and PA spoofing detection on the ASVspoof 2019 dataset using ensemble models. Below we summarise our contributions.

\begin{itemize}
    \item We build our models by discarding data points (Section~\ref{dataset_description}) ensuring non-overlap in spoofing conditions between training and validation for better generalisation.
    \item We demonstrate that combining information from deep and traditional machine learning approaches along with our dataset partition can improve model generalisation.
    \item We find that spoofed audio recordings for the PA task tend to have more silence at the end than bonafide recordings. We perform three different interventions proving that models exploit this fault in the dataset and achieve lower performance without these cues. 
    \item We make our dataset partition details and silence removal scripts available online\footnote{\scriptsize{\url{https://github.com/BhusanChettri/ASVspoof2019/}}}.
    
\end{itemize}

Our results suggest that performance metrics reported on the current PA dataset may be overestimating the actual performance of the models, which might become somewhat of a ``horse'' \cite{bobHorsePaper} that trivially sidesteps the actual problem, thus raising concerns about model validity as well as performance results. Prior work has addressed a similar issue of silence on the ASVspoof 2017 PA dataset~\cite{bhusanICASSP}, which calls for careful design and validation of the 2019 PA spoofing dataset\footnote{We have reported the ``silence'' issue to the challenge organisers.}.

\section{Task description and dataset} 
\label{dataset_description}

Given a speech recording the task is to build a spoofing countermeasure, a model, to automatically determine whether it is a bonafide (genuine) or spoofed, either generated through TTS, VC or a replayed recording. 

The ASVspoof 2019 LA and PA datasets were released as part of this year's challenge \cite{asvspoof2019_evaluationplan}. Both consist of $8$ male and $12$ female speakers in the training and development subsets. In LA, there are $2,580$ bonafide and $22,800$ spoofed utterances in the training set and $2,548$ bonafide and $22,296$ spoofed utterances in the development set. In PA, both the training and development sets has $5,400$ bonafide utterances, and $48,600$ and $24,300$ spoofed utterances in the training and development sets, respectively. The evaluation set has around $80,000$ and $135,000$ test utterances in the LA and PA datasets \cite{asvspoof2019overview,asvspoof2019_evaluationplan}.

The training and development subsets have similar spoofing algorithms/conditions in both the LA and PA datasets. We argue that using the same types of spoofing attacks during training and validation might lead to overfitting and poor generalisation on unseen attack conditions. Thus, we further partition the original training and development datasets for both LA and PA, ensuring non-overlap in spoofing attack conditions. We use a subset \textit{train\_{tr}} of the training set for model training and take a subset of the development (dev) set and partition it into \textit{dev\_{es}} for model validation and early stopping and \textit{dev\_{lr}} to build model ensembles through logistic regression.
Although taking these subsets requires discarding many samples, it allows us to test how well a model generalises to previously unseen attack conditions: 
The spoofing attack conditions and speakers in \textit{train\_{tr}} and \textit{dev\_{es}} are non-overlapping. In \textit{dev\_{lr}}, we use all spoofing conditions of the dev set but discard speakers that have been used in \textit{dev\_{es}}.


\section{Models in the Proposed Ensembles} \label{our_proposed_models}
In this section, we describe the approach used to design countermeasures for the LA and PA tasks of the ASVspoof 2019 challenge. A model ensemble is used in order to combine information from different countermeasure models employing various features and training procedures. This diversity leads to a powerful ensemble with good generalisation. 

\subsection{Deep models} 
We train five deep models using raw audio or time-frequency representations as input to minimise a binary cross-entropy~(CE) loss with an Adam optimiser and early stopping with a patience of $P$ epochs. As the dataset has more spoofed examples, we replicate the bonafide examples to ensure each batch contains an equal number of bonafide and spoofed examples, which helps stabilise the training. At inference time, we use the output layer sigmoid activation as a score. We provide model-specific training details below.


\subsubsection{Convolutional Neural Network (CNN)}
We use the CNN architecture from~\cite{bhusanSLT}, featuring 50\% dropout in the fully connected layers, a batch size of $32$, and a learning rate of $10^{-4}$. We train the model for $100$ epochs with an early stopping patience of $P=5$ and $P=2$ for the LA and PA tasks, respectively. We use an utterance-level mean-variance normalized log spectrogram\footnote{Power-spectrogram for the LA task and Mel-spectrogram with 80 mel bands (for computational reasons) for the PA task.}, computed using a 1024-point FFT with a hop size of 160 samples, as the input. For each task, we train two such CNN models, model A and B, on the first and last 4 seconds of each audio sample. We truncate or loop the spectrogram time frames to obtain a unified time representation.

\subsubsection{Convolutional Recurrent Neural Network (CRNN)}
We use a modified version of the CRNN architecture from our prior work \cite{Morfi:18a} (model C). We train the model for $500$ epochs with early stopping patience of $P=10$ for both the LA and PA tasks. As input, we use a mean-variance (computed on $train\_tr$ set) normalized log-Mel spectrogram of $40$ Mel bands, computed on the first 5 seconds of truncated or looped audio samples, using a 1024-point FFT with a hop size of 256 samples. During training, we use a batch size of $8$ and $32$ for the LA and PA tasks, respectively, with an initial learning rate of $10^{-5}$ that is halved on validation loss plateau with a patience of $P=5$ epochs, until $10^{-8}$.

\subsubsection{1D-Convolutional Neural Network} 
We use the network architecture from the sample-level 1D CNN \cite{lee2017sample} (model D). In total, the model consists of $9$ \textit{ReSE-2} blocks \cite{kim2018sample}. These blocks are a combination of \textit{ResNets} \cite{he2016deep} and \textit{SENets} \cite{hu2018squeeze}. We use the multi-level feature aggregation, where the outputs of the last three blocks are concatenated and followed by a fully connected layer of $1024$ units, batch normalization and ReLU layers, a 50\% dropout layer and a fully connected layer of $1$ unit with sigmoid activation. Each convolutional layer has filters of size $3$, \textit{L2} weight regularizer of $0.0005$ and all strides are of unit value. The raw audio input is $3.7$ seconds in duration and randomly sampled segments of this size are selected from the recordings. We loop shorter samples to obtain a unified time representation. We train the model using a batch size of $16$, learning rate of $10^{-4}$ and an early stopping patience of $P=25$ epochs.

\subsubsection{Wave-U-Net}
We use a modified version of the Wave-U-Net~\cite{Stoller2018a}, with five layers of stride four, and without upsampling blocks (model E). The outputs of the last convolution are max-pooled across time, reducing the parameter count and incorporating the intuition that the important features in the tasks are temporally local. Finally, we apply a fully connected layer with a single output to yield a classification probability. We train the model using a batch size of $64$, a learning rate of $10^{-5}$ and early stopping patience of $P=10$ for both the LA and PA tasks, where an epoch is defined as 500 update steps. To ensure the audio inputs have the same length, we pad all recordings with silence to $196608$ audio samples ($=12.23$ seconds). For the PA task, we also match real samples to their spoofed versions based on the speaker identity and utterance. We train on pairs of audio samples (discarding samples without any matches) and balanced batches, in order to stabilise the training process and improve generalisation by preventing the network from using speaker identity and utterance content for discrimination. 

\subsection{Shallow models} 
Additional to deep models, we use two different shallow \cite{fusion_birdspaper_salamon} models: Gaussian Mixture Models (GMMs) and Support Vector Machines (SVMs). 

\subsubsection{GMM}  \label{gmms}
We train three GMM models using $60$-dimensional static, delta and acceleration (SDA) mel frequency cepstral coefficients (MFCCs) \cite{mfccReference} (model F), inverted mel frequency cepstral coefficients (IMFCCs) \cite{imfccReference} (model G), and sub-band centroid magnitude coefficients (SCMC)  \cite{scmcReference} (model H), due to their performance on the ASVspoof 2015 and 2017 spoofing datasets \cite{sahidullah2015comparison, bhusanICASSP}. We use $128$ and $256$ mixture components for the LA and PA tasks respectively and train one GMM each for bonafide and spoof class. At test time, the score of each test utterance is obtained as the average log-likelihood ratio between the bonafide and spoof GMMs. We use the feature configuration from \cite{sahidullah2015comparison}.


\subsubsection{SVM} \label{svms}
We train two SVMs using i-vectors (model I) and the long-term-average-spectrum (LTAS) feature (model J) since they have shown good performance on prior spoofing datasets \cite{sahidullah2015comparison,hannah_TASLP2017,bhusanMLSP2018}. Inspired from \cite{secondbestSystemIS2015challenge} we fuse multiple i-vectors in our approach, each based on complimentary hand-engineered features, and manage to improve performance over a single i-vector based SVM. We train four different i-vector extractors using $60$-dimensional SDA MFCC, IMFCC, constant Q cepstral coefficients (CQCC) \cite{Todisco2016ANF} and SCMC features. We train the T matrix with $100$ total factors on both tasks and universal background model (UBM) with $128$ and $256$ mixtures on the LA and PA tasks, respectively and extract $4$ different $100$-dimensional i-vectors for every utterance. We use $400$-dimensional fused i-vectors for LA and $300$ for PA task. We perform mean-variance normalisation on the fused i-vectors and LTAS feature and train SVMs with a linear kernel and the default parameters of the Scikit-Learn \cite{scikit-learn} library. We train the UBM and T matrix using the MSR-Identity toolkit~\cite{msr2013sadjadi}.


\subsection{Ensemble models} \label{ensemble}
We define three ensemble models E1, E2 and E3 using the logistic regression implementation of the Bosaris \cite{bosaris2013brummer} toolkit. On the LA task, E1 combines models A, C through G and I, while E2 consists of A, B and G. On the PA task, E1 fuses all single models except D, and E2 combines models A through E. Finally, E3 combines models A and B on both LA and PA tasks.

\section{Experiments}
\label{sec:experiments}
\subsection{Experimental setup}
We train our models (single and ensemble) described in Section~\ref{our_proposed_models} using the \textit{train\_{tr}} and \textit{dev\_{lr}} sets respectively. We use \textit{dev\_{es}} for model validation, early stopping and hyper-parameter optimisation. We compare our models' performance with the baseline LFCC (model B1) and CQCC (model B2) feature based GMM models provided by the ASVspoof 2019 challenge organisers.


We evaluate our models using the minimum normalized tandem detection cost function (t-DCF) \cite{tomi_tDCF} metric, that takes both the ASV system and spoofing countermeasure errors into consideration, and is used as the primary evaluation metric in the ASVspoof 2019 challenge. We also evaluate our model performance independently with the equal error rate (EER) metric. Please refer \cite{asvspoof2019overview,tomi_tDCF} for details.

\subsection{Results}
\subsubsection{Development set} 
Table \ref{combined_dev_results} presents the results on the original development set for both LA and PA tasks. In general, the results suggest that PA task is harder than LA. For the PA task, our CNN performs noticeably better when operating on the last 4 seconds of audio (model B) instead of the first 4 seconds (model A), suggesting the presence of discriminative cues at the end of each audio signal which we confirm in Section \ref{analysing_models}. 
Furthermore, we observe a poor performance for models D and E. Apart from having to learn features directly from the raw audio, another reason could be that they involve zero-padding all signals or using a randomly selected audio segment for prediction, respectively, and thus might not be able to exploit such cues at the end of audio signals. 

\begin{table}
\caption{Results on the LA and PA development set. Bold: best performance, na: not applicable.}
\centering
\vspace{-1.5 mm}
\scalebox{1.0}{
\begin{tabular}{ccccc}
\toprule
\multirow{2}{*}{Model} &\multicolumn{2}{c}{LA} &\multicolumn{2}{c}{PA} \\
&t-DCF & EER\% &t-DCF & EER\% \\ 
\midrule
B1 & $0.0663$  &$2.71$ &$0.2554$  &$11.96$ \\
B2 &$0.0123$  &$0.43$  &$0.1953$  &$9.87$ \\
\midrule
A  &$0.0074$ &$0.32$ &$0.2795$ &$10.77$ \\
B &$0.0040$ &$0.27$ &$0.1672$ &$5.98$  \\
C &$0.1706$ &$5.65$ &$0.1223$ &$5.0$  \\
D &$0.36$ &$13.58$  &$0.9269$ &$36.28$  \\
E &$0.0745$ &$2.43$   &$0.4725$ &$21.16$  \\
F &$0.1805$ &$7.46$  &$0.2354$ &$10.88$  \\
G &$0.0438$ &$1.73$  &$0.2119$ & $8.94$  \\
H &$na$ &$na$ &$0.2787$ &$12.46$  \\
I &$0.0045$ &$0.16$ &$0.2537$ &$9.93$  \\
J &$na$ &$na$ &$0.3534$ &$13.6$   \\
\midrule
E1& \textbf{0.0} & \textbf{0.0} &  \textbf{0.0354} & \textbf{1.33}  \\
E2&$0.0002$ &$0.03$ &$0.0523$ &$1.85$  \\
E3&$0.0025$ &$0.2$ &$0.1316$ &$4.85$ \\
\bottomrule
\end{tabular}}
\label{combined_dev_results}
\end{table}

Our i-vector feature fusion approach (model I) shows impressive performance on the LA task but relatively poor performance on the PA task. One reason for this could be that the i-vectors extracted using hand-crafted features are not able to capture characteristics of unseen replay attack conditions. On both the LA and PA tasks, model G (IMFCC) outperforms model F (MFCC), suggesting that a focus on higher frequency information is beneficial as it might not be perfectly generated by the TTS and VC algorithms. Likewise, on the PA task, the playback device properties may impact high-frequency content. Finally, the poor performance of models H and J suggest that SCMC and LTAS features are not suitable for this task.

\begin{table}
\caption{Results on the LA and PA evaluation set. Bold, na: same as in Table \ref{combined_dev_results}.}
\centering
\vspace{-1.5 mm}
\begin{tabular}{ccccc}
\toprule
\multirow{2}{*}{Model} &\multicolumn{2}{c}{LA} & \multicolumn{2}{c}{PA} \\

&t-DCF & EER\% &t-DCF & EER\% \\ 
\midrule
B1 & $0.2116$  &$8.09$ &$0.3017$  &$13.54$\\
B2 &$0.2366$  &$9.57$ &$0.2454$  &$11.04$\\
\midrule
A &$0.1790$ &$7.66$ &$na$ &$na$ \\
B &$na$&$na$&$0.1577$ &$5.75$ \\
E1 &\textbf{0.0755} & \textbf{2.64} &$0.1492$ &$6.11$  \\
E2 &$0.2136$ &$9.57$ &$0.2913$ &$14.12$ \\
E3 &$0.2952$ &$10.63$ &\textbf{0.1465} & \textbf{5.43} \\
\bottomrule
\end{tabular}
\label{combined_eval_results}
\end{table}

As expected, our ensemble model appears to benefit from combining different models for both tasks, as indicated by the strong reduction in t-DCF and EER compared to all individual models. On both tasks, E1 performs better than E2 which in turn performs better than E3.

\subsubsection{Evaluation set}
Table \ref{combined_eval_results} shows the results on the evaluation set\footnote{Computed by the ASVspoof 2019 challenge organisers.}. On the LA task, model E1 has an EER of $2.64$\% and a t-DCF of $0.0755$, outperforming the baselines by a large margin and securing the third rank in the ASVspoof 2019 challenge. The superior performance of E1 over E2 and E3 suggests that fusing multiple models employing different features does provide complementary information useful for spoofing detection.

However, on the PA tasks our single model B outperforms ensemble models E1 (on the EER) and E2 (both metrics). Furthermore, our two model ensemble E2 (A+B) outperforms the five deep model ensemble E2 and nine model ensemble E1 reaching the lowest t-DCF of $0.1465$ and an EER of $5.43$\%. While these results suggest good model generalisation, it raises questions about the relevance of the cues used by model B as it is only trained on the last 4 seconds of each recording. 
Besides the poor performance of models D and E, the inferior performance of ensemble models on the evaluation set compared to the development set (Table \ref{combined_dev_results}) could be explained by model C making random predictions on the evaluation data (due to a bug we found after the challenge submission), but not on the development set -- which is corroborated by the fact that model C receives the second highest weight by logistic regression in both E1 and E2.

\begin{table}
\caption{Intervention (Int) results on the development set of \textbf{PA tasks}. Numbers to the left of arrow indicates performance without any intervention.}
\centering
\vspace{-1.5 mm}
\begin{tabular}{ccccc}
\toprule
Int &Model &t-DCF & EER\%  \\
\midrule
\multirow{3}{*}{I}& M1 &$0.2036 \rightarrow 0.2741 $ &$9.18 \rightarrow 13.27$ \\
& M2 &$0.1971 \rightarrow 0.2959 $ &$10.06 \rightarrow 15.59$ \\
& B &$0.1672 \rightarrow 0.5018$ &$5.98 \rightarrow 19.8$ \\
\midrule
\multirow{3}{*}{II}& M1 &$0.2036 \rightarrow 0.9528$ &$9.18 \rightarrow 54.76$ \\
& M2 &$0.1971 \rightarrow 0.9463$ &$10.06 \rightarrow 57.98$ \\
& B &$0.1672\rightarrow0.2626  $ &$5.98\rightarrow11.20$ \\
\midrule
\multirow{3}{*}{III}& M1 &$0.2036 \rightarrow 0.8614$ &$9.18 \rightarrow 41.09$ \\
& M2 &$0.1971 \rightarrow 0.9448$ &$10.06 \rightarrow 58.71$ \\
& B &$0.1672 \rightarrow 0.3129$ &$5.98 \rightarrow 12.85$ \\
\bottomrule
\end{tabular}
\label{intervention_results}
\end{table}

\section{Interventions on the PA task} \label{analysing_models}
In Table \ref{combined_dev_results} we find that for the PA task, the same CNN performs much better when trained on the last $4$ seconds of audio (model B) than on the first $4$ seconds (model A). We thus analyse a set of audio recordings for the PA task that were confidently classified by model B and find that spoofed audio tend to have more silence (zero-valued samples) at the end than bonafide examples. In comparison, silence at the beginning of the recordings is often shorter and does not appear to follow this pattern. Therefore, we hypothesize that any model (deep or shallow) trained on the PA dataset that does not specifically discard this information could exploit the duration of silence as a discriminative cue. This leads to countermeasure models that are easily manipulated, simply by removing silence from the spoofed signals to make the model misclassify them as a bonafide signal, and vice versa. To demonstrate this effect in practice, we perform three interventions on model B and the adapted\footnote{We use $128$ mixtures to train the LFCC (M1) and CQCC (M2) GMMs in contrast to $512$ mixtures used in the baselines B1 and B2.} baselines M1 and M2 by manipulating the silence at the end of the audio signal.

\subsection{Intervention I}
In this intervention we train the models on the original recordings with the silence but remove them during testing\footnote{We use a naive approach of counting the first consecutive zeros as silence and remove them.}. In Table~\ref{intervention_results}, a strong increase can be noticed in both EER and t-DCF for all models, suggesting that they indeed rely on the silence parts for prediction. We find that model B is most sensitive to this intervention, with t-DCF and EER rising by $0.3346$ and an absolute $13.82$\%, respectively. This could be due to deep models focusing more strongly on silences than the GMM models, which are trained on individual spectral frames and aggregate the score through averaging frame-wise likelihoods.


\subsection{Intervention II}
Here, we train the model with silence parts removed, but test on the original test recordings (with silence). The stable performance of the CNN (model B) over the GMMs in Table~\ref{intervention_results} suggests that the former is more robust against variations in silence duration.
On the other hand, we find a dramatic increase in error rates for M1 and M2. One interpretation for this is that bonafide and spoof GMM may assign a low likelihood to silence frames as they have not seen them during training. Thus, silence frames do not make large contributions to the final score making the task much harder.

\subsection{Intervention III}
In this intervention, we remove silence during training and testing to ensure that the audio samples do not share an easily exploitable cue. This forces the models to learn about the actually relevant factors of interest and thus provides more realistic performance estimates  (Table~\ref{intervention_results}). As in intervention II, model B shows a stable performance indicating good generalisation and discrimination capabilities. Models M1 and M2 on the other hand achieve a poor performance, possibly since their bonafide GMM models assign a high likelihood to spoofed frames as they are very similar to bonafide ones when only considering the speech frames.

\section{Discussion and conclusion}
\label{sec:conclusion}
In this paper, we approach the logical access (TTS and VC) and physical access (replay) spoofing detection problem on the ASVspoof 2019 dataset using ensemble models, demonstrating that combining models trained on different feature representations can be effective in detecting unseen spoofing attacks. We achieve good performance on the PA and $3^{rd}$ ranking on the LA tasks of the challenge. The PA task seems generally more difficult and should thus be the primary focus of future work.

Our intervention experiments in Section \ref{analysing_models} suggests that many models trained on the PA dataset can become somewhat of a ``horse'', where solving the actual problem is unintentionally avoided by exploiting silence as trivial cues. As the evaluation set also contains such silences, the reported performance metrics in this task currently overestimate the actual performance.
In addition to removing silence from the end of recordings, we also removed it from the beginning, but found that it has much less impact on performance and therefore do not report the results in this paper. However, due to our simple approach at silence removal, near-silent segments and silences between words within the recording might remain and could also serve as an undesirable discriminative cue and so should be investigated in future work.

We aim to perform further analysis on our deep models once the test set labels are released to the public, including the impact of the faulty deep model that produced random predictions on the evaluation set. 

\bibliographystyle{IEEEtran}
\bibliography{mybib}
\end{document}